\documentclass[12pt]{iopart}

\usepackage{amsfonts} 
\usepackage{graphicx}
\usepackage{textcomp,gensymb}
\usepackage{threeparttable}
\usepackage{booktabs,multirow}
\usepackage{natbib}
\usepackage[T1]{fontenc} 
\usepackage{setspace}
\begin{document}

\title{Transformable Reflective Telescope for optical testing and education}

\author{Woojin Park$^1$, Soojong Pak$^{1,*}$, Geon Hee Kim$^2$, Sunwoo Lee$^1$, Seunghyuk Chang$^3$, Sanghyuk Kim$^4$, Byeongjoon Jeong$^2$, Trenton James Brendel$^5$, and Dae Wook Kim$^5$}

\address{$^1$ School of Space Research and Institute of Natural Science, Kyung Hee University, Yongin 17104, Republic of Korea}
\address{$^2$ Korea Basic Science Institute, 169-148, Daejeon 34133, Republic of Korea}
\address{$^3$ Center for Integrated Smart Sensors, Korea Advanced Institute of Science and Technology (KAIST), Daejeon 34141, Republic of Korea}
\address{$^4$ Korea Astronomy and Space Science Institute, Daejeon 34055, Republic of Korea}
\address{$^5$ James C. Wyant College of Optical Sciences, University of Arizona, Tucson, Arizona 85721, USA}

\eads{\mailto{woojinpark@khu.ac.kr}, \mailto{soojong@khu.ac.kr, $^*$Corresponding author}}
\vspace{10pt}
\begin{indented}
\item[]July 2020
\end{indented}

\doublespacing

\begin{abstract}
We propose and experimentally demonstrate the Transformable Reflective Telescope (TRT) Kit for educational purposes and for performing various optical tests with a single kit. The TRT Kit is a portable optical bench setup suitable for interferometry, spectroscopy, measuring stray light, and developing adaptive optics, among other uses. Supplementary modules may be integrated easily thanks to the modular design of the TRT Kit. The Kit consists of five units; a primary mirror module, a secondary mirror module, a mounting base module, a baffle module, and an alignment module. Precise alignment and focusing are achieved using a precision optical rail on the alignment module. The TRT Kit transforms into three telescope configurations: Newtonian, Cassegrain, and Gregorian. Students change telescope configurations by exchanging the secondary mirror. The portable design and the aluminum primary mirror of the TRT Kit enable students to perform experiments in various environments. The minimized baffle design utilizes commercial telescope tubes, allowing users to look directly into the optical system while suppressing stray light down to $\sim$10$^{-8}$ point source transmittance (PST). The TRT Kit was tested using a point source and field images. Point source measurement of the Newtonian telescope configuration resulted in an 80\% encircled energy diameter (EED) of 23.8 $\mu$m.
\end{abstract}

\maketitle

\section{Introduction}
\label{sec:intro}
Reflective telescopes are primarily based on Newtonian, Cassegrain, and Gregorian designs. The Newtonian optical design is the simplest of the three and is generally well-adapted for commercial telescopes. Optical antennas and reflective objective mirrors use modified Cassegrain or Gregorian optical systems \citep{jiang2015,gaivoronskii2012}. Recently, Gregorian designs have become more important in large-aperture telescopes, as the Gregorian secondary mirror can be optically conjugated to the upper layer of the Earth's atmosphere enabling the use of adaptive optics (AO) techniques for diffraction-limited observation \citep{alexander2003}. For example, the aplanatic Gregorian optical design of the Giant Magellan Telescope (GMT) utilizes seven segmented primary mirrors, each 8.4 meters in diameter, and adaptive secondary mirrors \citep{johns2014}. 

Newtonian, Cassegrain, and Gregorian optical systems rely chiefly upon a primary mirror and a secondary mirror. The parabolic primary mirror is used to correct for spherical aberration is common across all three telescope designs. The principal differentiating factors among these three optical designs are the location and surface shape of the secondary mirror. The Transformable Reflective Telescope (TRT) may be reconfigured to each of the optical designs by simply changing the secondary mirror.

A few transformable kits were introduced which are exchangeable for Newtonian or Cassegrain design. One setup uses a 45$\degree$ - folding mirror within a Cassegrain design to change the setup into a Newtonian telescope \citep{jppi1}. In this kit, the Newtonian setup requires a folding mirror which is larger in size than the secondary mirror found in the Cassegrain design. The other kit places the secondary mirror along the optical axis of the primary mirror. By sliding the convex hyperbolic secondary mirror along the optical axis to the primary mirror, the Cassegrain design is converted to a Newtonian telescope \citep{jppi2}. There is also a method that includes two-optical systems within a single setup by installing the Newtonian and Cassegrain secondary mirrors on a rotatable mount \citep{fr2923303}. The Gregorian configuration, however, utilizes a much longer focal length than the other designs, making it technically difficult to implement all configurations in a single setup.

The transformable optical system also requires optical alignment and baffling to suppress stray light. This necessitates the use of an alignment module, baffles, and optomechanical structures. We present a flexible astronomy educational kit which can be configured into each of the previously described reflective telescopes. 

Optical design of the transformable optical system is discussed in section \ref{sec:systemDesign}.\ref{sec:OpticalDesign} and \ref{sec:systemDesign}.\ref{sec:tol}. Optomechanical design and structural analysis results are presented in section \ref{sec:systemDesign}.\ref{sec:design}. Fabrication process and mirror surface quality will be explored in section \ref{sec:fab}.\ref{sec:mirrorfab}. Section \ref{sec:fab}.\ref{sec:align} describes the optical alignment procedure for this system. Stray light consideration and baffle design are examined in section \ref{sec:strayCG}. Finally, section \ref{sec:perform} presents optical performance measurements of the TRT Kit.

\section{TRT system design and analysis}
\label{sec:systemDesign}
\subsection{Flexible TRT optical design}
\label{sec:OpticalDesign}

\begin{figure}[ht]
\centering\includegraphics[width=10cm]{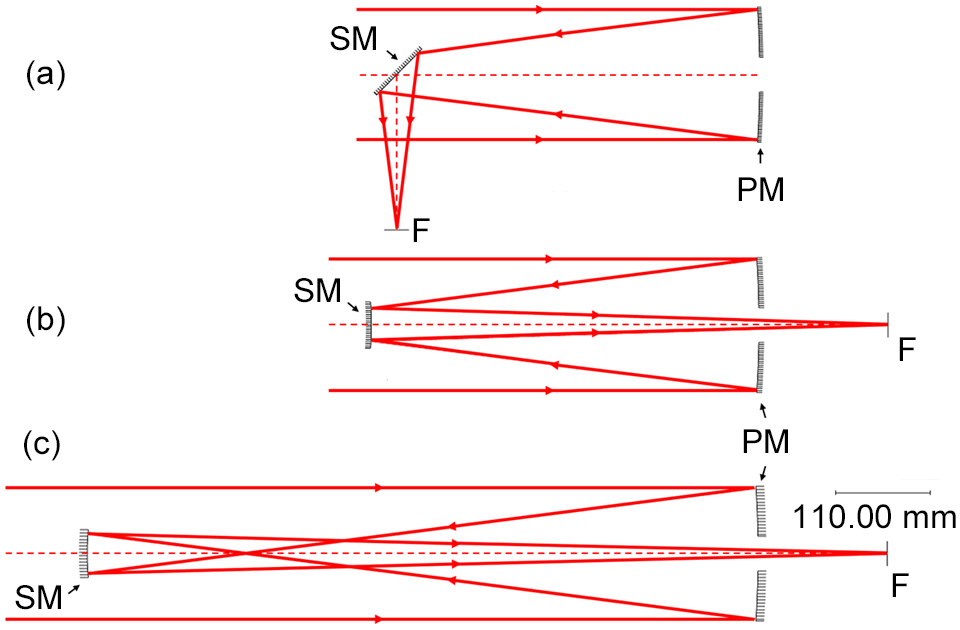}
\caption{\label{opticdesign}Optical layout of  (a) Newtonian, (b) Cassegrain, and (c) Gregorian telescope. (PM: primary mirror, SM: secondary mirror, F: focal point).}
\end{figure}

The TRT Kit transforms into three traditional two-mirror optical systems, the Newtonian, Cassegrain, and Gregorian telescopes. Figure~\ref{opticdesign} illustrates the telescope configurations. In each case, incident rays are reflected from the paraboloidal primary mirror and the secondary mirror sequentially, so that rays converge to a single focal point as depicted in the figure. The Newtonian, Cassegrain, and Gregorian systems use flat, convex hyperboloidal, and concave ellipsoidal secondary mirrors, respectively. The optical design parameters of the primary and secondary mirrors used for the TRT Kit are summarized in Table~\ref{table:OpticalSpec}.

\begin{table*}[ht]
\centering
\begin{threeparttable}
\caption{Optical design parameters of the TRT Kit.}
\footnotesize
\begin{tabular}{l| l| c| c| c}
\toprule
\multicolumn{2}{c|}{} & Newtonian & Cassegrain & Gregorian \\
\hline
\centering
\multirow{3}*{Primary mirror}  & Aperture (mm) & \multicolumn{3}{c}{150} \\
                                                  & Focal ratio & \multicolumn{3}{c}{F/4} \\
                                                  & Surface type & \multicolumn{3}{c}{Paraboloid} \\\hline
\multirow{3}*{Secondary mirror} & Aperture (mm) & 50 & 50 & 50 \\
                                                & Surface type & Flat  & Hyperboloid & Ellipsoid \\
                                                & Conic constant & -  & -2.778 &  -0.458 \\\hline
\multicolumn{2}{l|}{Effective focal length (mm)} & 600 & 2400 & 3115 \\
\multicolumn{2}{l|}{FOV\tnote{a}~ (with the 22.2 $\times$ 14.8 mm sensor){ } ($^\prime$)} & 127.2 $\times$ 84.8 & 31.8 $\times$ 21.2 & 24.5 $\times$ 16.3 \\
\bottomrule
\end{tabular}
\begin{tablenotes}
\item[a]{\footnotesize Fields of view}
\end{tablenotes}
\label{table:OpticalSpec}
\end{threeparttable}
\end{table*}
\raggedbottom

All secondary mirrors have a 50 mm clear-aperture. The radius of curvature of the Cassegrain and Gregorian secondary mirrors are -400 mm and +300 mm, respectively. The axial distance between the primary and secondary mirrors, which is also called despace, are 450 mm for Newtonian and Cassegrain, and 778.9 mm for Gregorian. The back focal length of Newtonian, Cassegrain, and Gregorian are 178.5 mm, 600 mm, and 928.9 mm, respectively. The conic constants for the TRT Kit Cassegrain and Gregorian configurations are -2.778 and -0.458, respectively. It is worth to note that optimal conic constants for the primary and secondary mirrors can be selected to desensitize the telescope decentration tolerance for two-mirror systems \citep{lucimara2013}. For the TRT Kit, however, parabolic primary mirror is required for the Newtonian configuration and such a conic constant optimization is not available.

In the proposed TRT Kit, three optical systems share a single primary mirror. Reconfiguring the telescope system requires a simple exchange of the secondary mirror while the rest of the optomechanical parts remain fixed.

\subsection{Tolerance analysis}
\label{sec:tol}
\begin{figure}[!ht]
\centering\includegraphics[width=10cm]{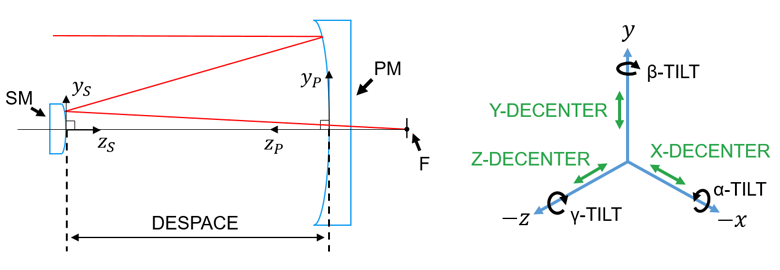}
\caption{\label{coordinate}The coordinate system for tolerance analysis (same abbreviations with Figure~\ref{opticdesign}).}
\end{figure}

Optical performance changes from assembly and alignment errors are expected based on tolerance analysis. We used CODE V and ZEMAX for sensitivity analysis and Monte-Carlo simulation, respectively. Figure~\ref{coordinate} shows the coordinate system for tolerance analysis. 

\begin{figure*}[!ht]
\centering\includegraphics[width=15.5cm]{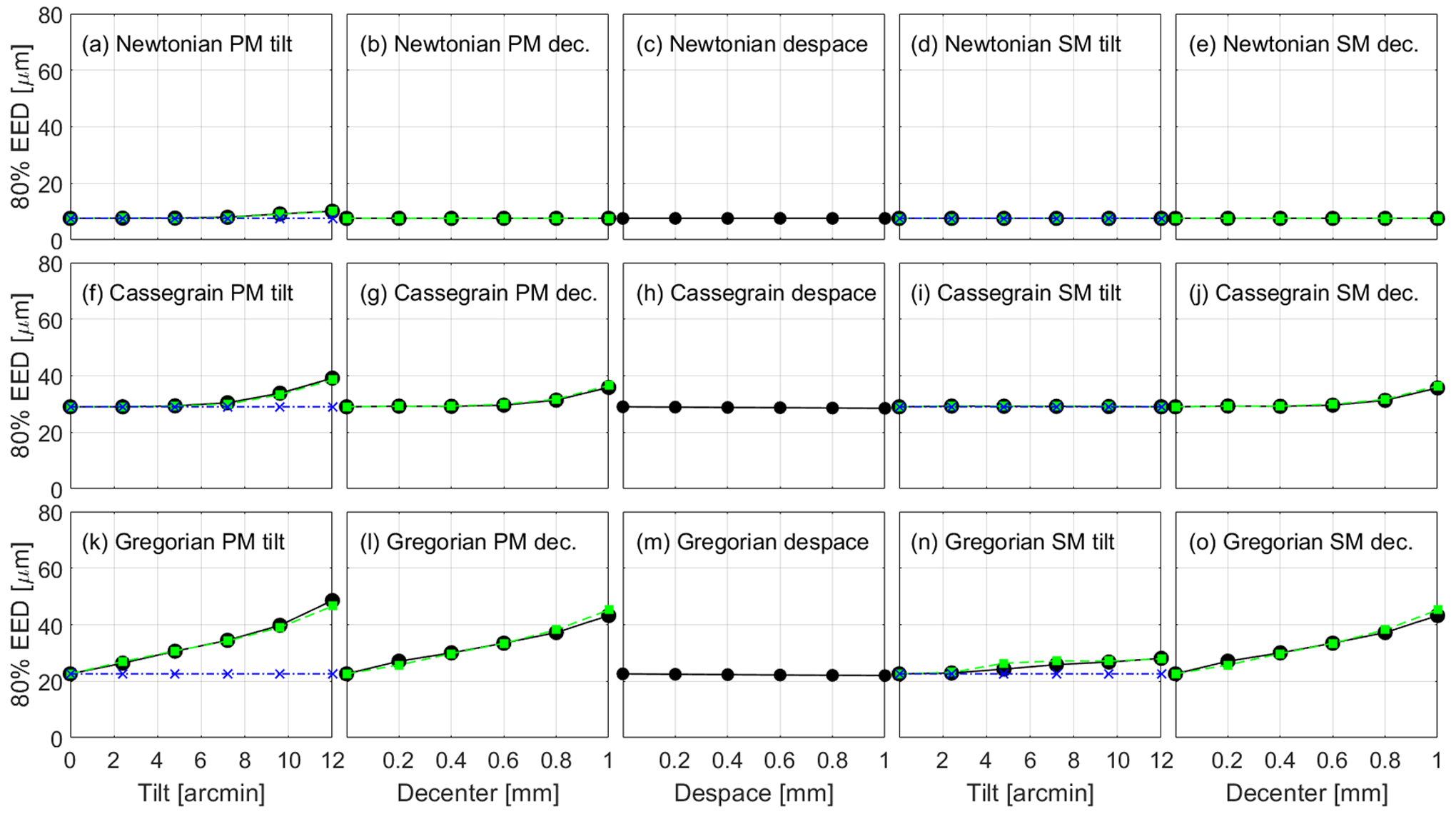}
\caption{\label{Sensitivity}Sensitivity analysis of Newtonian (top), Cassegrain (middle), and Gregorian (bottom) systems: (a, d, f, i, k, n) $\alpha$- (circle), $\beta$- (square), and $\gamma$- (cross) tilts, (b, e, g, j, l, o) x- (circle) and y- (square) decenters, and (c, h, m) despace.}
\end{figure*}

Sensitivity analysis is completed to gauge optical performance degradation by optical alignment errors and mechanical deformations. Surface $\alpha$-, $\beta$-, and $\gamma$- tilts, x- and y- decentration of each mirror, and despace are examined for sensitivity in the Newtonian, Cassegrain, and Gregorian telescope configurations. The criterion of the analysis is 80\% encircled energy diameter (EED) for the 550 nm wavelength, and the focus is set as the compensator. Figure~\ref{Sensitivity} shows sensitivity analysis results for the image center. 80\% EED is displayed for each of the perturbed system parameters across all three telescope configurations. Calculated sensitivities for negative perturbations are symmetric to the positive cases.

The Newtonian secondary mirror is a simple flat used to fold the optical path and does not affect optical performance. It is shown as constant 80\% EED curves across decenters of the primary and secondary mirrors, tilts of the secondary mirror, and despace in Figure~\ref{Sensitivity}. Despace is not critical to optical performance for all three systems. Decenters of the primary and secondary mirrors have the same trends because they are directly coupled to each other. $\alpha$- and $\beta$- tilts of the primary mirror are the most sensitive parameters for all three systems.  

We performed Monte-Carlo simulations as a statistical tolerance analysis to examine the effects of all errors simultaneously \citep{park2020}. Monte-Carlo simulation is performed for $\alpha$- and $\beta$- tilts, x- and y- decenters, despace, and surface irregularity. The simulation is evaluated with 5000 trials at the 550 nm reference wavelength. The focus is set as the compensator to recover performance. The performance criterion is 12 $\mu$m for 80\% EED which is based on the Nyquist sampling requirement for a sensor with with 6 $\mu$m pixels.

Tabel~\ref{table:monte} lists the final tolerance ranges of three TRT configurations from the Monte-Carlo simulation. The Gregorian is the most sensitive of the three systems analyzed. Overall tolerance ranges are acceptable and are within general fabrication and alignment errors. Alignment accuracy of the TRT Kit will be discussed in Section~\ref{sec:fab}.

\begin{table}[ht]
\centering
\caption{Tolerance limits of Newtonian, Cassegrain, and Gregorian from the Monte-Carlo simulations.}
\medskip
\begin{threeparttable}
\footnotesize
\begin{tabular}{l| l| c| c| c}
\toprule
\multicolumn{2}{l|}{Parameter}                       & Newtonian & Cassegrain & Gregorian \\
\hline
\centering
\multirow{2}*{$\alpha$-, $\beta$- Tilt }        & PM & $\pm$ 23$^\prime$  & $\pm$ 4$^\prime$     & $\pm$ 2.5$^\prime$ \\
                                                & SM & $\pm$ 32$^\prime$   & $\pm$ 10$^\prime$    & $\pm$ 5$^\prime$ \\\hline
\multicolumn{2}{l|}{x-, y- Decenter }                & $\pm$ 1 mm   & $\pm$ 0.6 mm   & $\pm$ 0.5 mm\\
\multicolumn{2}{l|}{Despace}                        & $\pm$ 3 mm   & $\pm$ 1 mm    & $\pm$ 0.7 mm \\
\multicolumn{2}{l|}{Surf. Irregularity }       & $\pm$ 2.5 $\lambda$ & $\pm$ 0.50 $\lambda$   & $\pm$ 0.35 $\lambda$ \\\hline
\multicolumn{2}{l|}{Focus\tnote{a}}                      &\multicolumn{3}{c}{$\pm$ 20 mm} \\
\bottomrule
\end{tabular}
\begin{tablenotes}
\item[a]{\footnotesize The focus is set as the compensator.}
\end{tablenotes}
\label{table:monte}
\end{threeparttable}
\end{table}
\raggedbottom

\subsection{Optomechanical Design of TRT}
\label{sec:design}
\begin{figure}[!ht]
\centering\includegraphics[width=10cm]{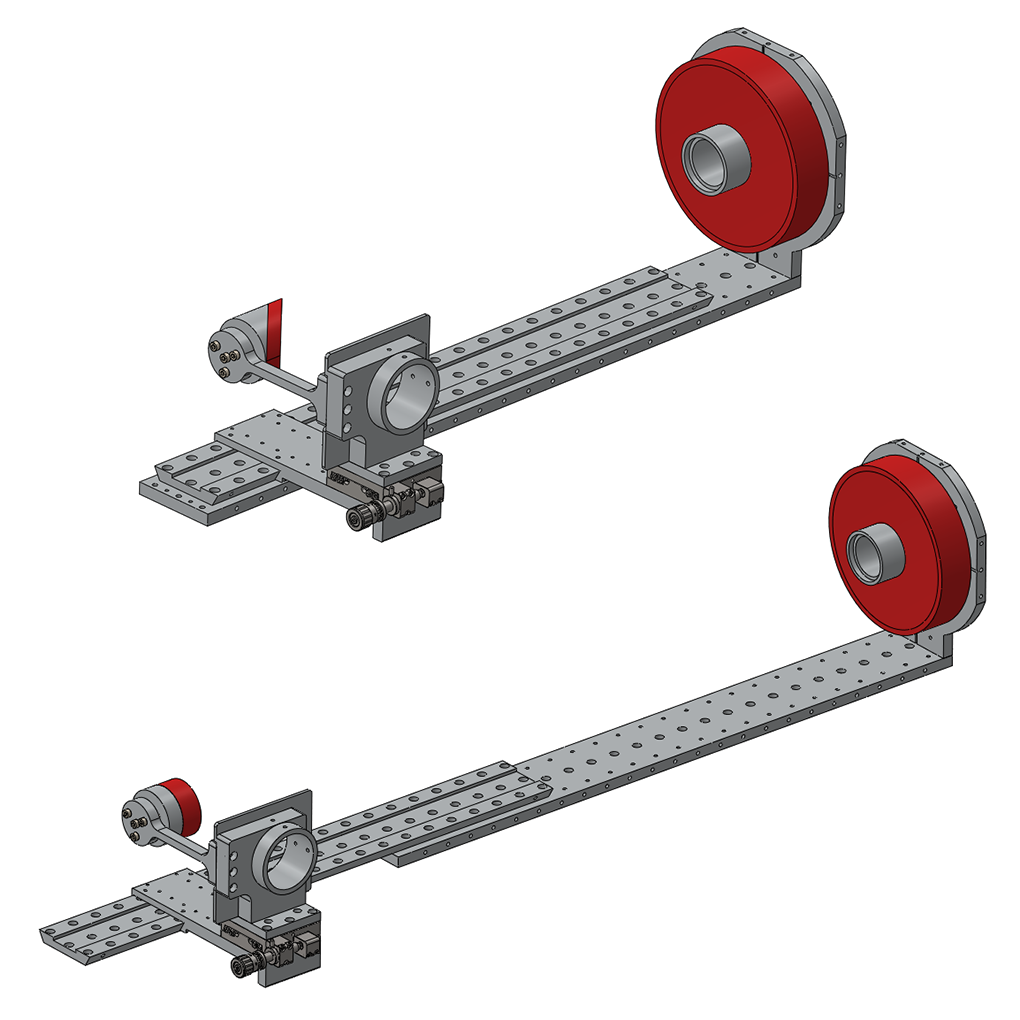}
\caption{\label{threeTRT}3D optomechanical modeling of the TRT Kit. Newtonian (top) and Gregorian (bottom) systems are shown.}
\end{figure}

The TRT Kit is a modularized and portable system, enabling easy assembly of each type of telescope and adaptable optical test setup. The total dimensions are 610 mm (L) $\times$ 158 mm (W) $\times$ 188 mm (H), and the total weight is around 2 kg. The length extends to 970 mm in the Gregorian configuration. Illustrations for the Newtonian and Gregorian configurations are displayed in Figure~\ref{threeTRT}. The mounting base module contains the array of holes and taps like an optical table so the TRT Kit becomes transformable, and the equally spacing holes can be used for measuring some distance such as the focal length and despace. A precision linear stage is used for fine focusing in the accuracy of >10 $\mu$m, and 2-inch camera adapters are found near the two focal points. Red parts indicate optical mirrors. 

\begin{figure}[!ht]
\centering\includegraphics[width=10cm]{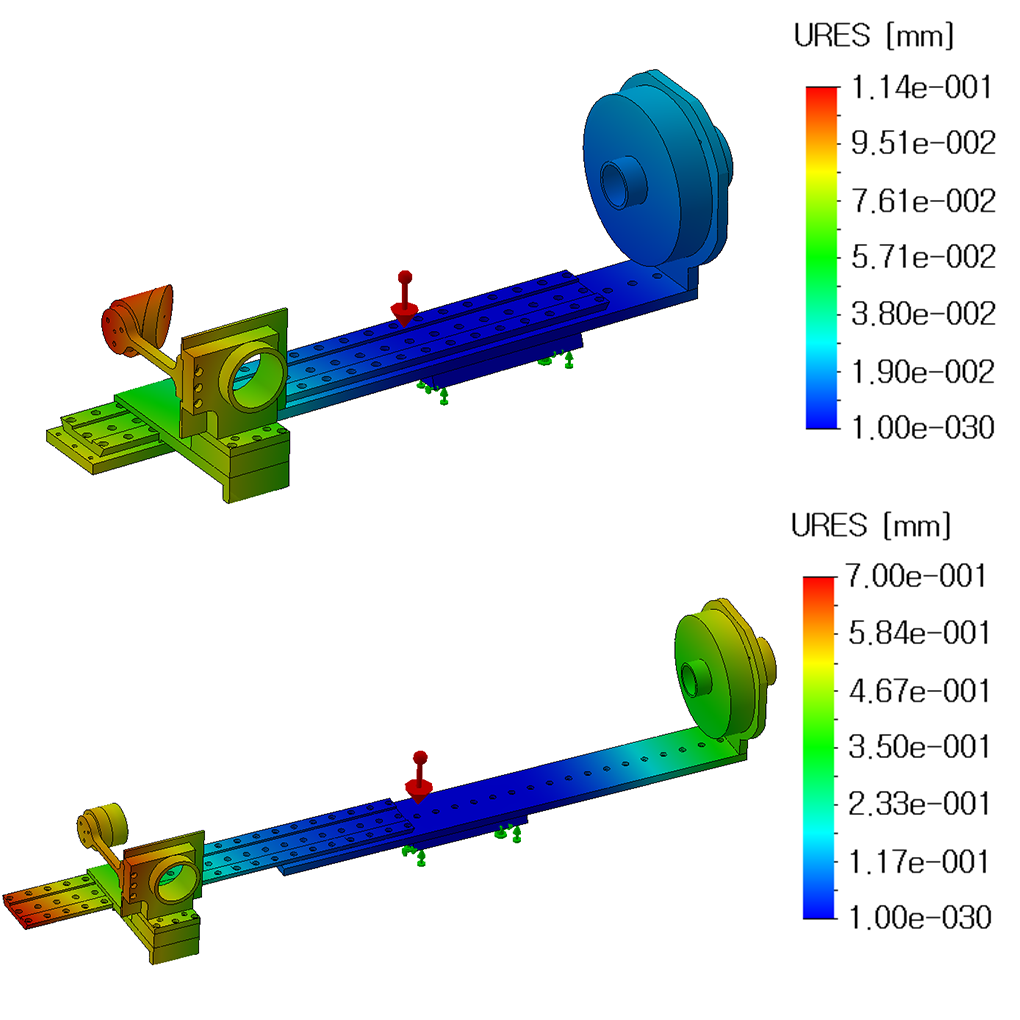}
\caption{\label{fea}Static analysis results of the compact mode (top) and the extended mode (bottom). The red arrows represent gravity, and fixed (reference) positions are indicated with green arrows.}
\end{figure}

Static analysis of the TRT Kit was accomplished with SolidWorks (Dassault Systems) mechanical design software. Optomechanical structures are made of aluminum alloy (Al6061-T6). The boundary condition used in the simulation fixes the bottom of the dovetail (green arrows in Figure~\ref{fea}) in place, allowing other parts to move relative to this static mount. 

The maximum mechanical deformation of optomechanical structures due to its weight is 0.11 mm and 0.7 mm for the compact (Newtonian, Cassegrain) and extended (Gregorian) configurations (Figure~\ref{fea}). Based on tolerance analysis results, the mechanical deformations are not critical to optical performance.

\section{TRT manufacturing, assembly, and alignment}
\label{sec:fab}
\subsection{Aluminum mirror fabrication}
\label{sec:mirrorfab}
The primary mirror is made of aluminum alloy (Al6061-T6) with protected aluminum coated on the mirror surface. The aluminum mirrors have advantages over other substrates in handling and thermal stability, characteristics required for the portability of the system \citep{hadjimichael2002}. The TRT primary mirrors are easy to replace with mirrors of different aperture size, surface type, and material. Fabrication processes of aluminum mirrors are as follows: pre-fabrication, stress relieving and aging, single-point diamond turning (SPDT), and protective coating. The entrance pupil diameter of the primary mirror is 150 mm and it has an F-number of 4 (see, Table~\ref{table:OpticalSpec}). 

The SPDT (Freeform 700A; Precitech) process produces a high-quality surface figure on the primary mirror. We employ a compensation strategy  \citep{zhang2015} to minimize Low-Frequency Error (LFE) \citep{kim2015} and Mid-Frequency Error (MFE) of the mirror surface. 

We used the Ultrahigh Accurate 3-D Profilometer (UA3P, Panasonic Corp.) to mechanically measure the mirror surface shape error using contact profilometry instead of optically measuring the errors \citep{chkhalo2016,wyant1972,li2008,yu2015,lylova2015}.

\begin{figure}[!ht]
\centering\includegraphics[width=10.0cm]{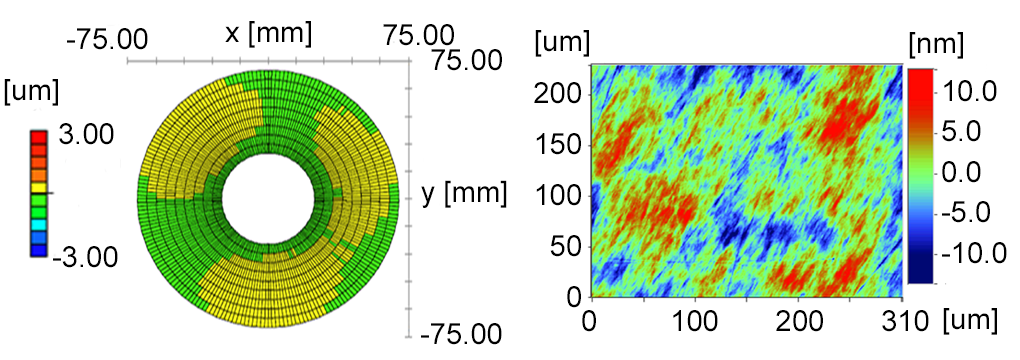}
\caption{\label{mirror}Surface measurement results of the primary mirror. The overall surface shape is measured with the UA3P (left). The surface micro-roughness is measured with NT2000 (right).}
\end{figure}

Figure~\ref{mirror} represents surface measurement results of the fabricated aluminum primary mirror. The surface figure error of the mirror is 0.067 $\mu$m RMS (Root Mean Square) and 1.84 $\mu$m PV (Peak to Valley). The average surface roughness (Ra) is 4.8 nm as measured with the WYKO NT2000 vertical scanning and phase-shifting interference microscope. 

The primary and secondary mirrors are aluminum coated with >95\% reflectivity, so the total throughput of the two mirror telescopes are about 90.3\% for all three configurations.

\begin{figure*}[ht!]
\centering\includegraphics[width=16cm]{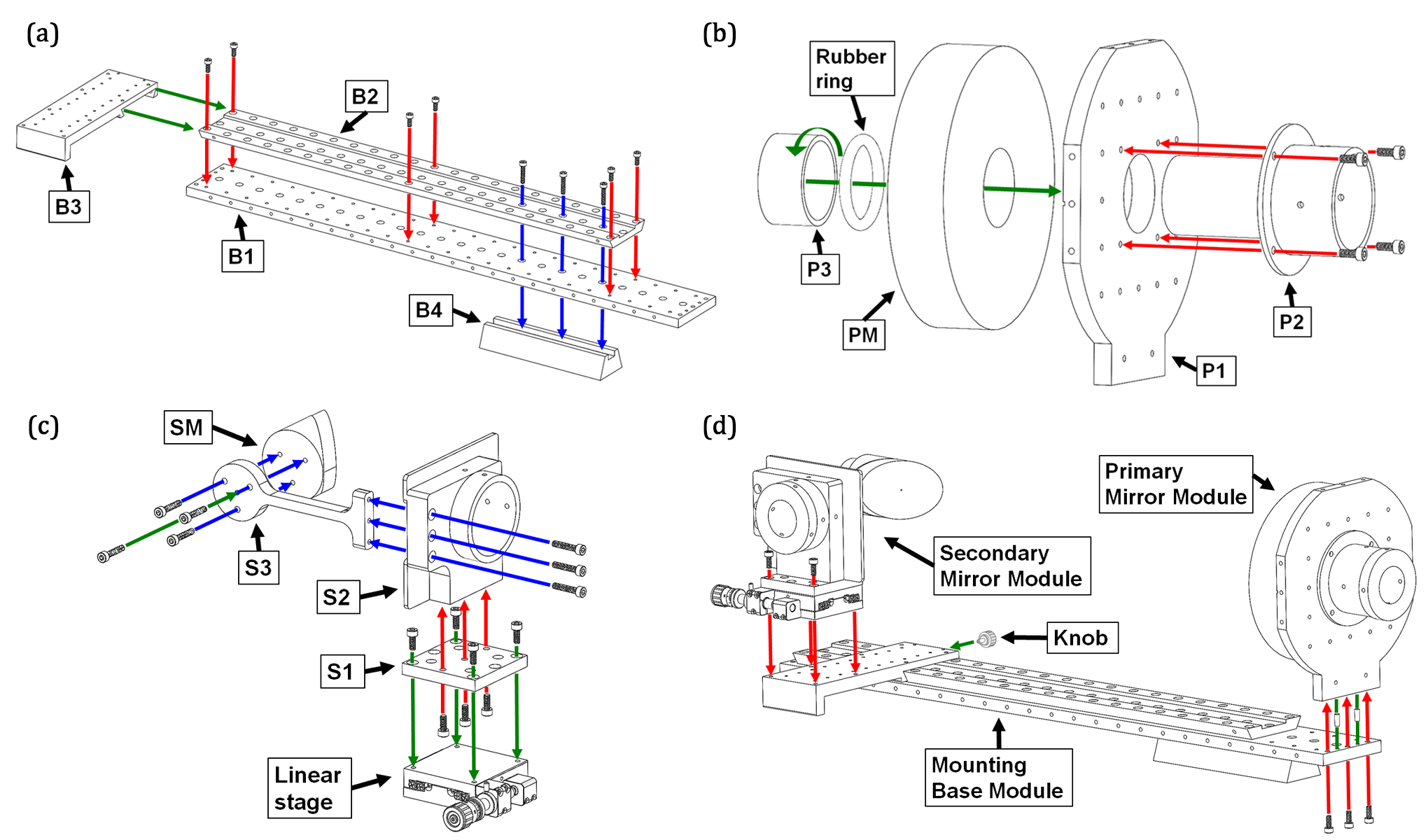}
\caption{\label{assem}The TRT assembly, including (a) the mounting base module, (b) the primary mirror module, (c) the secondary mirror module, and (d) all three modules assembled in the final TRT Kit configuration. Colored arrows indicate the screw and pin attachment points in the assembly process.}
\end{figure*}
\subsection{Assembly method}
\label{sec:assemble}
The TRT Kit has three main modules, a primary mirror module, a secondary mirror module, and a mounting base module. The modules are designed for interchangeability. The total number of optical and optomechanical components is 13. Screws and pins maintain the optical alignment. Because the modules are completely independent of one another, students may assemble each module separately and then cooperate to construct the final version of the TRT Kit. 

Figure~\ref{assem} illustrates the assembly method of each module. (a) The mounting base module consists of four mechanical parts, labeled B1 to B4, which support the primary and secondary mirror modules. The TRT Kit mounts onto commercial telescope mounts, thanks to a universal dovetail adapter (B4) attached to the base. (b) The primary mirror is mounted to P1 by threading P3 to P2, and a rubber ring is inserted between P3 and the primary mirror to protect the reflective mirror surface. (c) The tip and tilt of the secondary mirror is adjusted with screws attached to the mirror through S3 and spaced 120$\degree$ apart for three-point alignment. The linear stage for precise mirror transition is attached to S1, S1 is attached to S2, and S2 is attached to S3, the secondary mirror mount. The linear stage is then attached to B3 on the mounting base for precise positioning of the secondary mirror relative to the primary mirror. (d) The primary mirror module is fastened to the mounting base module with pins connecting P1 to B1. This mount configuration guarantees that the primary mirror support is attached perpendicular to the optical axis, drastically improving the simplicity of the design because only the secondary mirror must be aligned to the fixed primary mirror.

The primary mirror is aligned within $\pm$ 1.2$^\prime$ tolerance, which is measured by using Coordinate Measurement Machines (CMM), thanks to the pins between P1 and B1. The secondary mirror is replaced to other types using screws for S3 and the secondary mirror. The mounting base module needs to be reassembled for the Gregorian system (see, the bottom panel in Figure~\ref{threeTRT}).

\subsection{Optical alignment process}
\label{sec:align}
Shack-Hartmann wave-front sensors and CMM are commonly used for optical alignment \citep{wu2016}. Three-Point Laser Alignment (TPLA) is another common practice used to align optics. TPLA uses three mounted lasers aligned parallel to the optical axis of the primary mirror. 

\begin{figure}[!ht]
\centering\includegraphics[width=10cm]{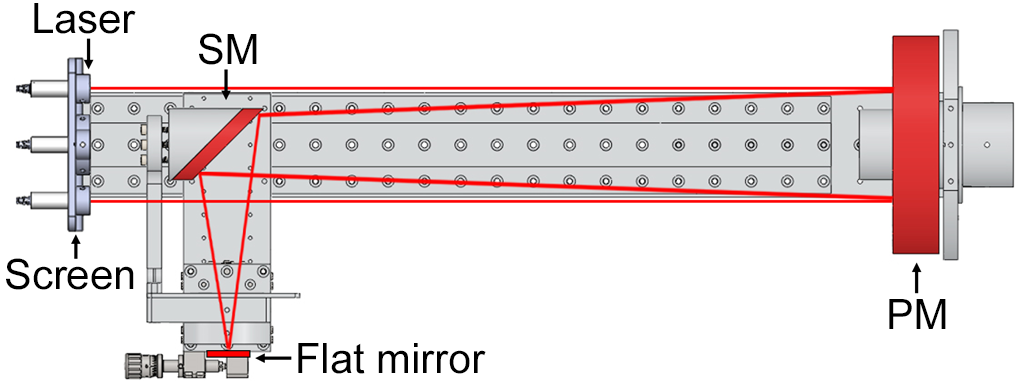}
\caption{\label{align}Layouts of TPLA for the Newtonian telescope. This method can be adapted to the other types of telescopes (same abbreviations with Figure~\ref{opticdesign}).}
\end{figure}

The laser mount contains a screen with holes at points optically conjugate to the sources. A flat mirror is placed at the focal plane of the optical system. First, the laser sources are emitted at the screen. Then, the beams reflect off of the primary mirror and converge towards the secondary mirror. For the Newtonian design, the flat secondary mirror then folds the optical path. Finally, the beams reflect off of the flat at the focal point and return through the optical system to the screen. The laser sources reflect back through the optical system to the conjugate points of the sources only when all optical components are well aligned (see, Figure~\ref{align}). 

\begin{figure}[!ht]
\centering\includegraphics[width=12cm]{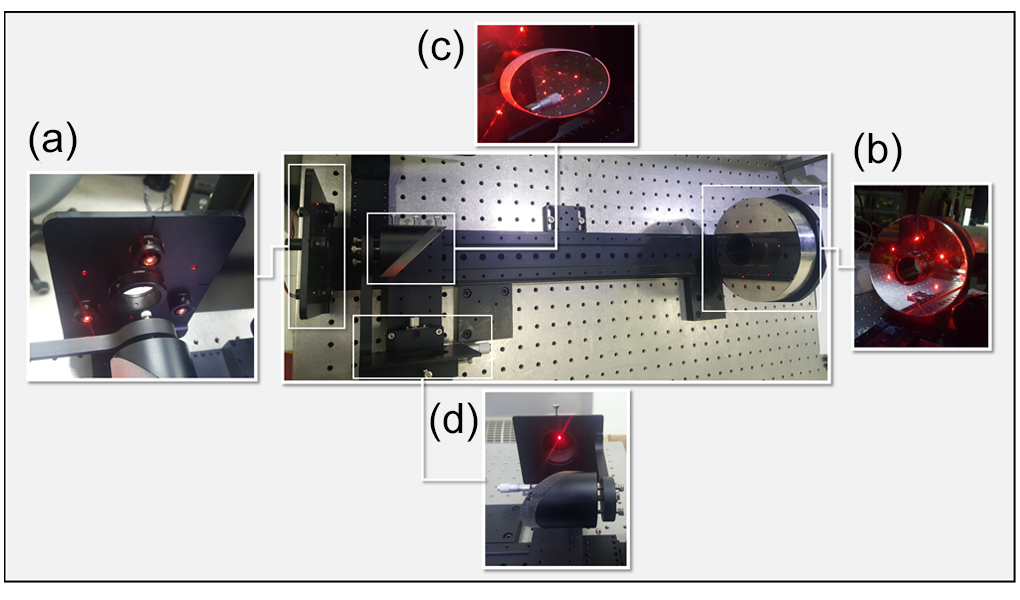}
\caption{\label{alignpic}The TRT Kit alignment with TPLA method. (Center) The laser mount attached to the TRT Kit. (Sub-pictures) (a) The lasers come from the source, (b) and are reflected by the primary mirror, (c) the secondary mirror, (d) and the flat mirror. After traveling back through the system, (a) the beams reach the screen.}
\end{figure}

Figure~\ref{alignpic} shows pictures of TPLA alignment process. There are 3 mm diameter alignment holes at the conjugate points of the laser sources (panel (a) in Figure~\ref{alignpic}) that are used for fine alignments. We measure coincidence of the laser point and the hole by eyes within 0.5 mm uncertainty that corresponds to 0.71$^\prime$, 0.18$^\prime$, and 0.14$^\prime$ tilt errors of the secondary mirror in Newtonian, Cassegrain, and Gregorian, respectively. These alignment accuracy are acceptable by comparing to tolerances of the telescopes (Table~\ref{table:monte}).

\section{TRT baffle and stray light control}
\label{sec:strayCG}
Stray light analysis and effective baffle design are necessary to suppress unwanted light that degrades the image. Stray light is extraneous, unwanted light which is detected by the sensor reflection from mechanical structures or scattering from optical components. Stray light presents itself as noise in the image, reducing the signal to noise ratio (SNR).

The baffle design of on-axis reflective optical systems, such as Cassegrain, and Ritchey-Chretien telescopes is a well-defined process described by mathematical models. An iterative method is needed to design proper baffle systems \citep{ho2009,kumar2013,song2002}.

\begin{figure}[!ht]
\centering\includegraphics[width=10cm]{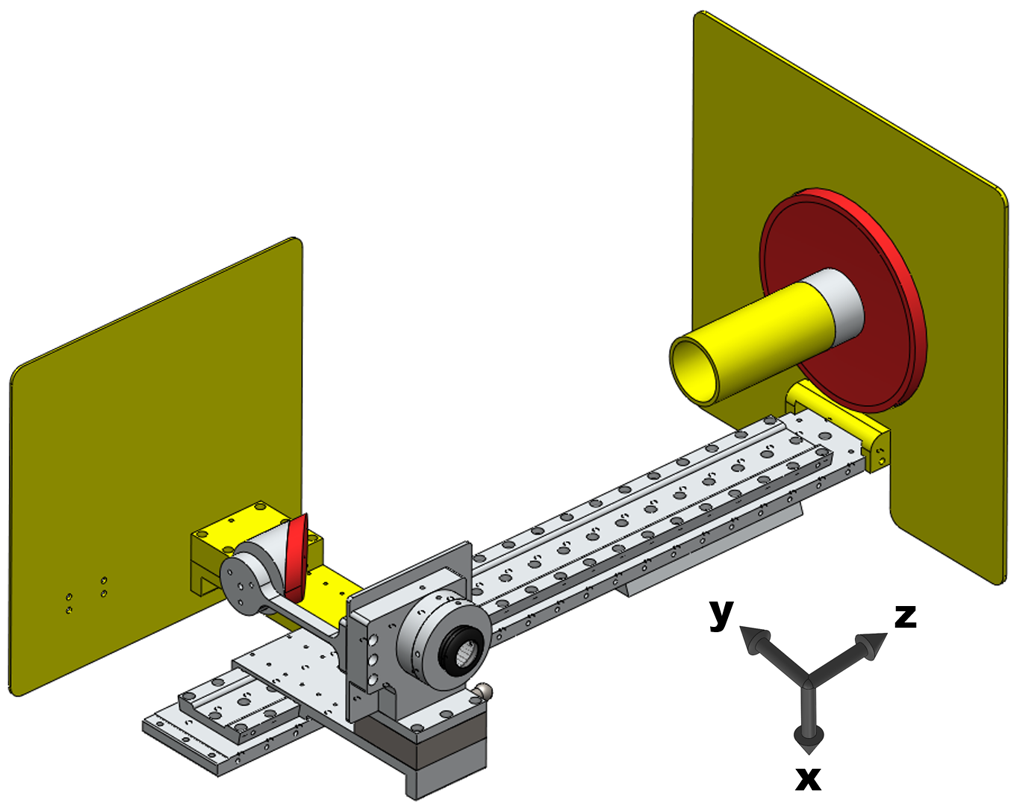}
\caption{\label{baffle}The TRT Kit with baffle structures. Yellow colors represent the baffles of the telescope.}
\end{figure}

For this portable optical device, stray light suppression is critical and even more important than it is for instruments operated indoors, as there are more potential sources of stray light when imaging outdoors. Baffle structures are easy to assemble with screws and T-mounts. Figure~\ref{baffle} demonstrates baffle structures suitable for all three optical configurations, though only the Newtonian configuration is shown. The plate baffles behind the primary and secondary mirrors are designed to effectively suppress critical stray light paths that directly illuminate the detector. These critical ray paths are of particular importance for the folded Newtonian system, which is susceptible to more potential sources of stray light as a result of the folded design.

\begin{figure}[!ht]
\centering\includegraphics[width=10cm]{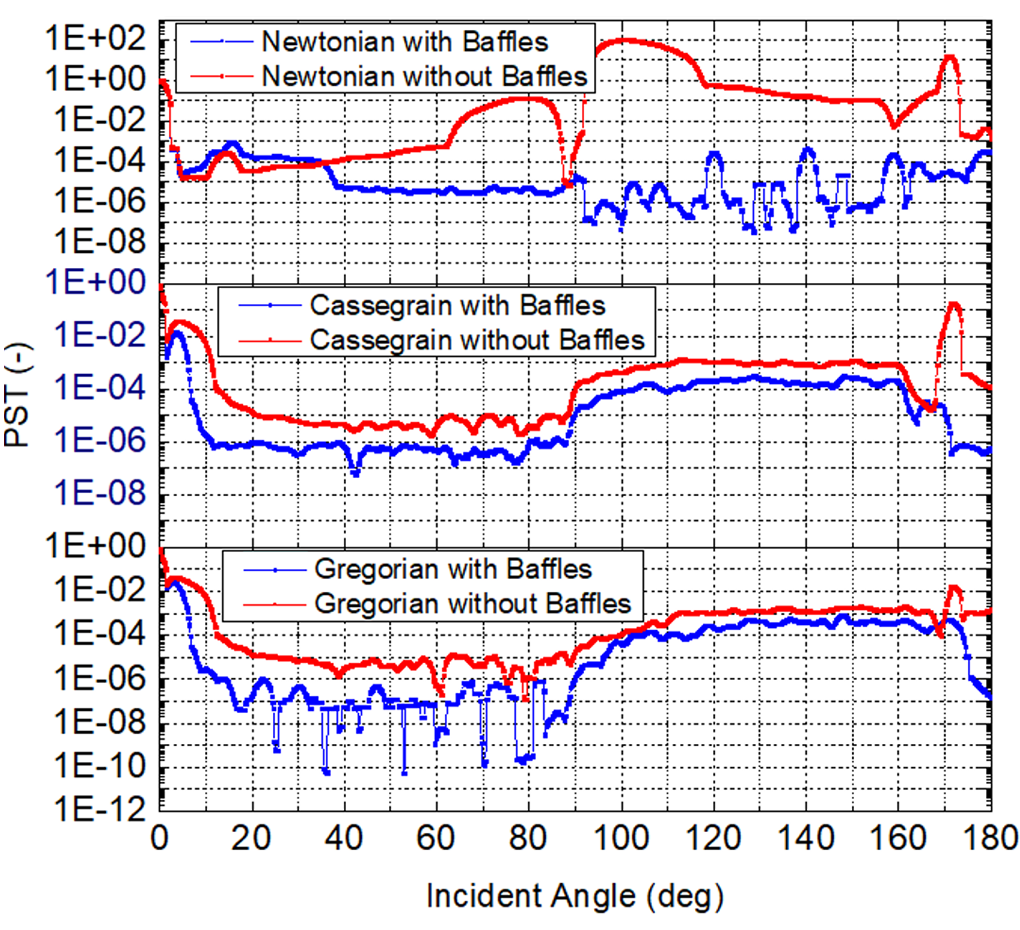}
\caption{\label{stray}Stray light analysis results of (a) Newtonian, (b) Cassegrain, and (c) Gregorian systems. Blue dots represent the PST with baffles, and red dots indicate the PST without baffles.}
\end{figure}

Stray light analysis was completed with LightTools software (Synopsys Inc.). In the simulation, a 6500 K black body source was used to illuminate the optical system, approximating the spectral distribution of daylight. Assuming ideal diffusive surfaces, baffles and optomechanical structures scatter light with 1$\%$ and 5$\%$ Lambertian reflectance, respectively. Incidence angles of simulated stray light paths span 0$\degree$ - 180$\degree$ for all systems.
The amount of stray light suppression can be defined with the Point Source Transmittance (PST):
\begin{eqnarray}
PST(\theta,\lambda_\textrm{0}) = \frac{L_\textrm{D}(\theta,\lambda_\textrm{0})}{L_\textrm{A}(\theta,\lambda_\textrm{0})}
\end{eqnarray}
where $L_\textrm{D}$($\theta$,$\lambda_\textrm{0}$) is spectral radiance on the detector, and $L_\textrm{A}$($\theta$,$\lambda_\textrm{0}$) is spectral radiance on the entrance aperture.
The PST represents how much stray light can be suppressed by the baffle structure as a function of incidence angle. Figure~\ref{stray} presents results from stray light analysis for the TRT Kit. Blue dots represent the PST with baffles, and red dots indicate the PST without baffles. The incident angles of 0$\degree$, 90$\degree$, and 180$\degree$ indicate the incident light along +z, -y, and -z direction (see, Figure~\ref{baffle} for the coordinate system).  

Baffles suppress 75\% - 95\% of stray light from all incidence angle (0$\degree$ to 180$\degree$). The PST reaches a minimum of around 10$^{-8}$ with baffles. Note that some of sharp drops in PST near the 20$\degree$ to 90$\degree$ incidence angles for the Gregorian system are a result of under-sampling.

Compared to the other systems, the detector is most exposed to stray light without the baffle in the Newtonian system. Incoming stray light from specific angle (95$\degree$ - 118$\degree$, and around 170$\degree$) causes substantial degradation of image quality without the use of baffles. Suppression of stray light for incidence angles between 40$\degree$ \ - \ 180$\degree$ is particularly significant for the Newtonian system with the addition of baffles. 90\% to 99.99\% of stray light is suppressed in this region.

\section{TRT system performance and application}
\label{sec:perform}
\subsection{System performance}
\begin{figure}[!ht]
\centering\includegraphics[width=15cm]{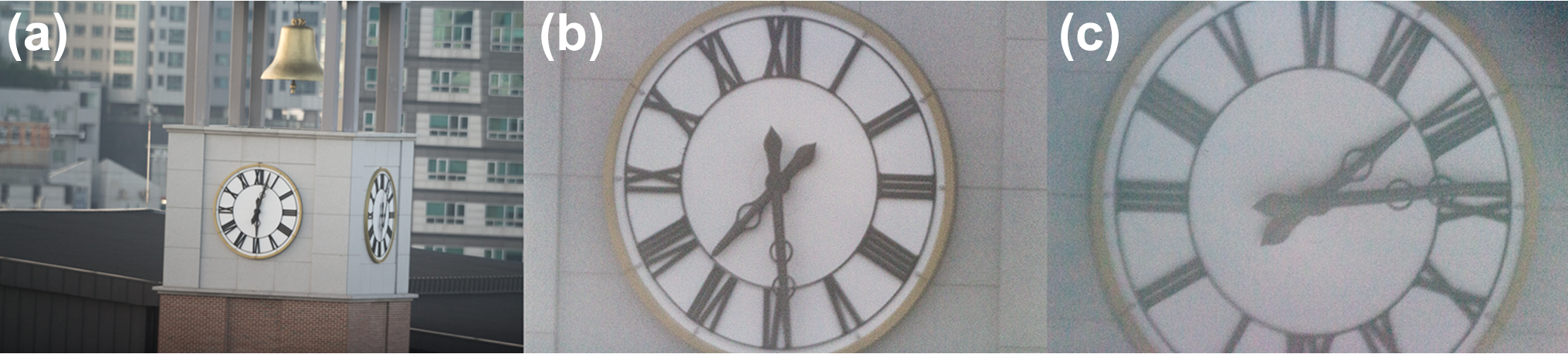}
\caption{\label{fov}The TRT Kit field test pictures. The pictures were captured with the (a) Newtonian, (b) Cassegrain, and (c) Gregorian telescopes, respectively. Each telescope has a different FOV, and the picture with the Gregorian telescope is reversed.}
\end{figure}

The optical performance of the TRT Kit was evaluated with three optical tests: observations of field images, point source tests, and night sky observations. Figure~\ref{fov} includes pictures that are taken with (a) the Newtonian, (b) Cassegrain, and (c) Gregorian telescopes. These three pictures clearly indicate the difference in FOV of the three configurations (see, Table~\ref{table:OpticalSpec}). The picture captured with the Gregorian telescope also shows the optically reversed image. The focal length can be derived from FOV of the observed image by calculating the plate scale.

\begin{figure}[!ht]
\centering\includegraphics[width=10cm]{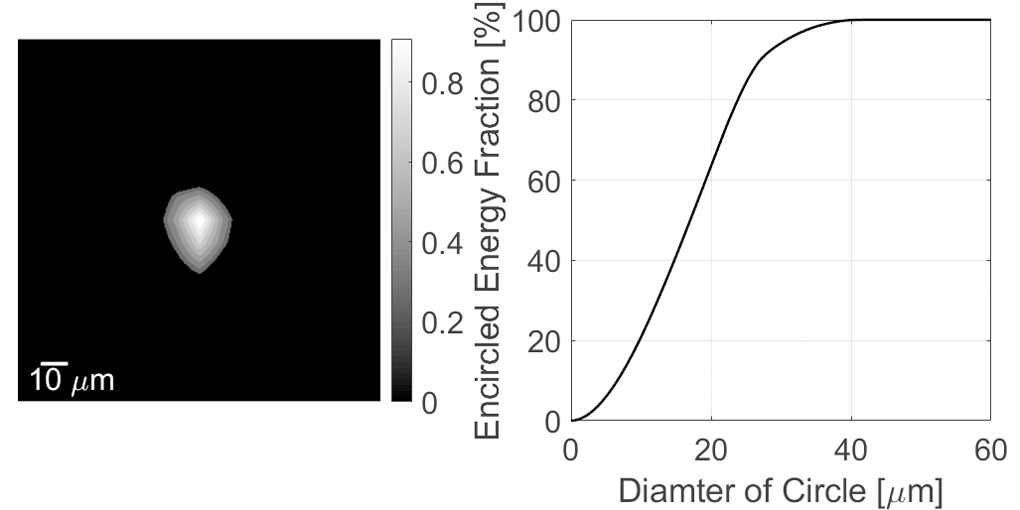}
\caption{\label{measSpot}The point source image (left) and EED (right).}
\end{figure}

Point source tests were also performed using the Newtonian configuration at the on-axis. Figure~\ref{measSpot} shows the point source image with 23.8 $\mu$m 80$\%$ EED, corresponding to $\sim$3.7-pixels on a binned APS-C sensor. This is an appropriate spot size for scientific measurements because the Nyquist sampling theorem is satisfied, thereby avoiding under-sampling. 

\begin{figure}[!ht]
\centering\includegraphics[width=12cm]{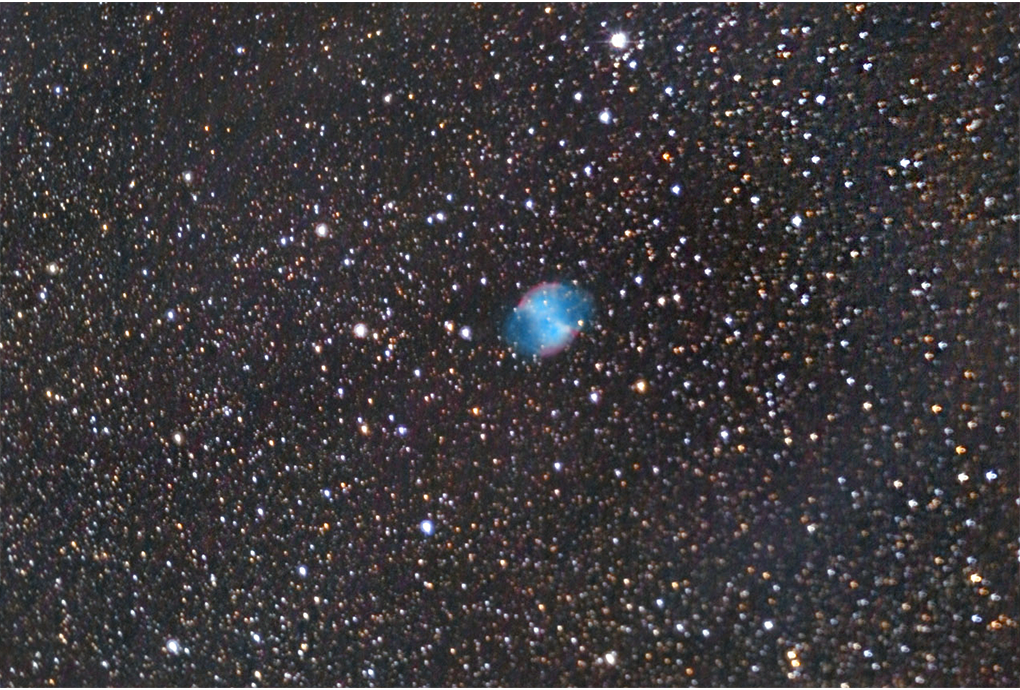}
\caption{\label{m27}The M27 nebula with stars that were observed using the Newtonian telescope.}
\end{figure}

Figure~\ref{m27} depicts the Messier 27 (M27) nebula captured using the Newtonian TRT Kit configuration. The observation demonstrates great optical performance capable of imaging detailed structures of the extended source and the fine circular shapes of stars across the field.

\subsection{Application to spectroscopy}
\begin{figure}[!ht]
\centering\includegraphics[width=10cm]{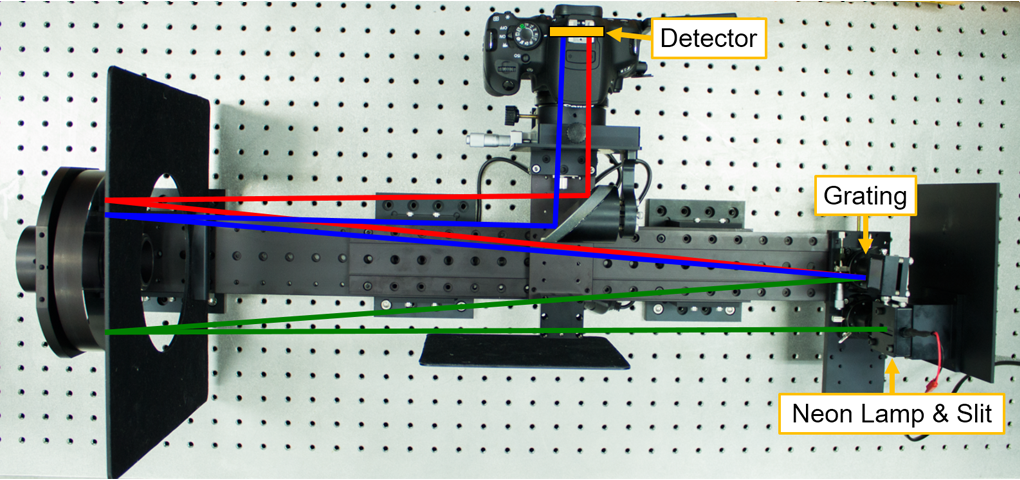}
\caption{\label{specmodule}The spectrometer module with the TRT Kit.}
\end{figure}

Since the TRT Kit is modularized, it transforms to the spectrometer by installing the spectrometer module, which consists of a grating, a slit, and a light source. Figure~\ref{specmodule} includes the main TRT Kit, the baffles, and the spectrometer module to configure the Ebert-Fastie spectrometer. 

\begin{figure}[!ht]
\centering\includegraphics[width=10cm]{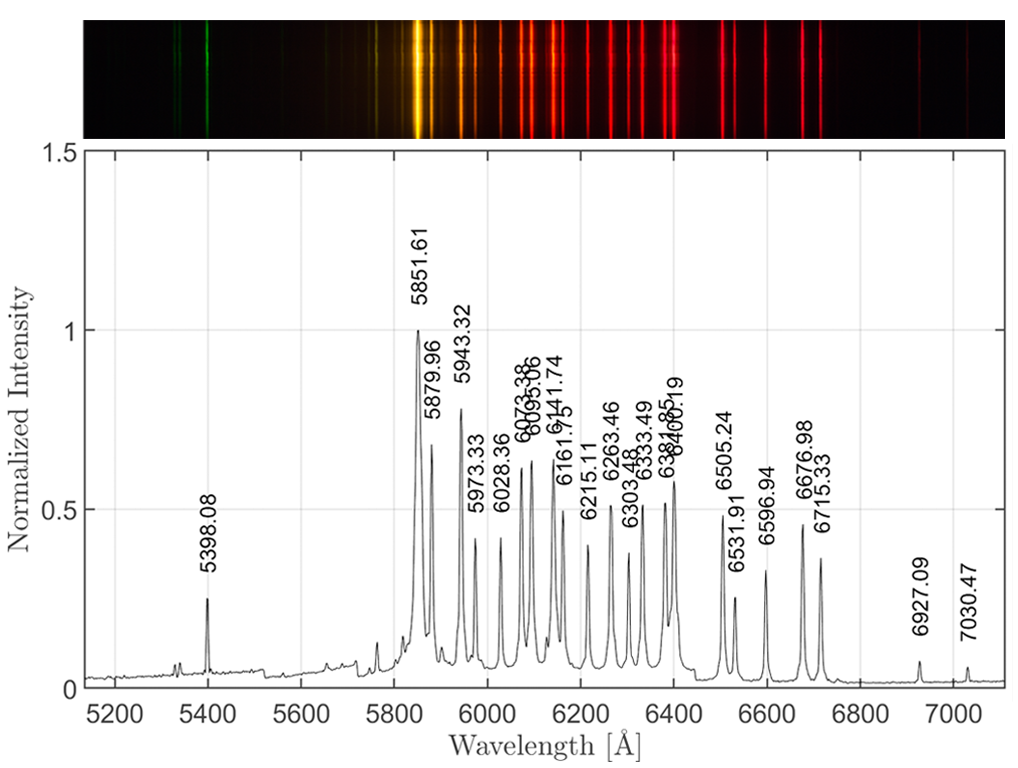}
\caption{\label{neonspec}Neon spectrum images from the TRT Kit and the spectrometer module.}
\end{figure}

A grating with 300 (grooves/mm) groove density and 4.3$\degree$ blaze angle, and a neon lamp were installed for the spectrum that is illustrated in Figure~\ref{neonspec}. At the position of the neon lamp, we can also place the focal plane of another telescope system to see the spectrum of distant targets. The commercial digital camera was installed for taking images. After the wavelength calibration, we clearly identified spectrum lines of the lamp in the wavelength range 510 nm to 720 nm.

\section{Discussion and summary}
\label{sec:discussion}
We developed the TRT Kit to transform into the Newtonian, Cassegrain, and Gregorian telescopes, as well as the Ebert-Fastie spectrometer. The modular structure of the TRT Kit maximizes versatility for various optical tests. Students only need to replace the secondary mirror to switch to other types of telescope or optical system.

The maximum optomechanical deformations by self-weight are 0.11 mm for Newtonian and Cassegrain configurations, and 0.7 mm for the Gregorian design. Even though these deformations may degrade the optical performance, the errors are acceptable by comparing to Monte-Carlo simulation results. 

Optimized baffle structures are designed for stray light suppression. It suppressed stray light from 10$^{-1}$ to 10$^{-8}$ PST across all angles of incidence.

The aluminum parabolic primary mirror was fabricated with SPDT. Surface errors on the primary mirror are $\le$ $\lambda$/8 RMS and the average surface roughness is 4.8 nm. The mirrors are made of aluminum to prevent imaging quality degradation by nonuniform thermal expansion or contraction.

The TRT Kit has a simple optical alignment procedure which requires only secondary mirrors to be aligned. The TPLA module makes optical alignment nearly effortless.

Point source measurement resulted in a 23.8 $\mu$m 80$\%$ EED. When observing the night sky, we were able to distinguish the fine structures of the M27 nebular. The TRT Kit is useful not only for optical experiments but also as an astronomical telescope. 

The TRT Kit is a versatile, portable telescope and optical test system. It can be utilized for many optical experiments involving spectroscopy, Gaussian theory, Fourier optics, and for developing an adaptive optics system by using the different types of telescope configurations such as Gregorian and Cassegrain, which create unique conjugate planes requiring different deformable mirror and wavefront sensor configurations. The compact size and portable design of the TRT Kit enable its use in many different environments. 

The TRT Kit transformable design concept has obtained a domestic patent from the Korean Intellectual Property Office (application number KR10-2015-0153977) but the commercialization has not yet been processed. Estimated fabrication cost of optomechanical parts is about 600 - 1000 US dollars which probably affordable for high school or university. The TRT Kit is suitable for both educational purposes and scientific research. We would like to widely distribute the transformable telescope kit to students, researchers, and others who are interested in using the Kit.  

\medskip
This work was supported by the Creative Convergence Research Project in the National Research Council of Science and Technology of Korea (CAP–15–01–KBSI). This work was a collaborative research work with the Wyant College of Optical Sciences. This research was made possible in part by the Technology Research Initiative Fund Optics/Imaging Program at the University of Arizona.

\newcommand{\newblock}{}
\bibliographystyle{aasjournal}
\bibliography{TRT_ref}

\end{document}